\documentstyle{elsart}

\input epsf

\begin{document}

\begin{frontmatter}

\title{Intercalibration of Cherenkov Telescopes in Telescope Arrays}

\author{Werner Hofmann}

\address{Max-Planck-Institut f\"ur Kernphysik, \\
D 69029 Heidelberg, P.O. Box 103980}

\begin{abstract}
A simple analysis technique is described which
allows to intercalibrate the response
of imaging atmospheric Cherenkov telescopes in stereoscopic telescope
arrays at a level of 1-2\%.
\end{abstract}

\end{frontmatter}

With the next-generation imaging atmospheric Cherenkov telescopes
currently under construction, stereoscopic arrays of telescopes
such as VERITAS \cite{veritas}, CANGAROO III \cite{cangaroo} and 
H.E.S.S. \cite{hess}
play a dominant role. Stereoscopic systems of Cherenkov telescopes
for VHE gamma-ray astrophysics  provide - compared to a single telescope -
superior angular resolution, energy resolution and 
background rejection, as demonstrated by the
HEGRA group \cite{hegra1,hegra2}. Making use of information concerning the
impact point of showers and the height of the shower maximum,
energy resolutions around 10\% are possible \cite{eres}. To actually
obtain this resolution, the relative response of the telescopes,
i.e. the relation between the light yield and the digital counts
provided by the ADC system must be known with very good precision.
Photon detectors within one telescope can be flat-fielded using
a light pulser in the center of the dish. Telescope-to-telescope
intercalibration is more complicated, and recently e.g. the use
of cosmic ray trigger rates and the so-called throughput
method has been advocated for this purpose \cite{throughput}. In the 
following, a simple technique for telescope intercalibration is
discussed (see also \cite{kruger}).

Once a telescope is properly flat-fielded, one needs only one global
calibration factor per telescope, relating the image size - the sum 
of the amplitudes of all image pixels - to the photon yield, up to
some overall calibration factor common to all telescopes of an array.
The response of two telescopes can be compared by selecting
showers with cores in the middle between two telescopes, and comparing
the image sizes of both telescopes. However, in systems with more
than two telescopes, one needs to be careful not to introduce a bias
due to other telescopes involved in the triggering or the event
reconstruction. Testing the technique with the HEGRA telescope
system, the following procedure was followed:
\begin{itemize}
\item To compare the response of two telescopes, events were selected
where both telescopes under consideration
had triggered (and possible others in addition)
\item Only information from the two telescopes under consideration
was used in reconstruction the shower; images from other telescopes
were ignored.
\item To reduce the influence of trigger thresholds, a cut well above 
the threshold was applied; the sum of the image sizes of the two
telescopes had to be above 200 photoelectrons.
\item To improve image quality and reconstruction, only images within
a radius of $1.75^\circ$ from the camera center were used, avoiding
image truncation. The (stereo) angle between the two image axes had
to be at least $30^\circ$. Reconstructed shower directions had to be
within $0.7^\circ$ from the telescope axis.
\end{itemize}
The relative response of two telescopes $i$ and $j$ was then characterized
by plotting the size asymmetry $a_s$ versus the asymmetry $a_r$ in
impact distance (Fig.~1), where
$$
a_{s(ij)} = {s_i - s_j \over s_i + s_j }
$$
and 
$$
a_{r(ij)} = {r_i - r_j \over r_i + r_j }
$$
Here, $s_i$ is the image size in telescope $i$ and $r_i$ its
distance to the reconstructed shower impact point, measured 
perpendicular to the shower axis. The asymmetry parameters
were chosen because they treat both telescopes in a symmetrical
fashion; one could also use $\log(s_i/s_j)$ or similar quantities.
\begin{figure}
\begin{center}
\mbox{
\epsfxsize10.0cm
\epsffile{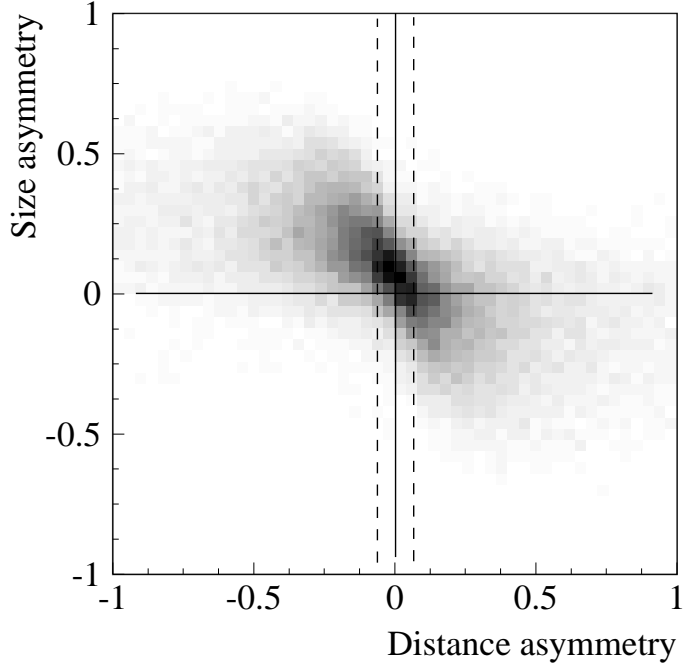}}
\end{center}
\caption{Distribution of size asymmetry $a_s$ versus distance asymmetry $a_r$,
for a telescope pair of the Mkn 501 data set. The dashed lines
indicate the cut range used to derive $a_s(a_r \approx 0)$.}
\label{fig_cal}
\end{figure}
To evaluate the relative
response of the two telescopes, one needs to determine the
average $a_s$ for $a_r = 0$. This can be achieved by averaging
$a_s$ for bins of $a_r$ and then fitting a smooth curve to
$<a_s(a_r)>$, or, if event statistics is high enough, by simply
cutting on $a_r$. For the following examples, the mean value
of $a_s(a_r \approx 0)$ were obtained by fitting a Gaussian
to the distribution of $a_s$ for $|a_r| < 0.05$. The method
was first applied to Monte-Carlo simulations of the HEGRA
telescope system in its 1997 configuration, with four telescopes;
one of the outer telescopes was still incomplete. In the 
simulation, the response of the four telescopes (for historical
reasons labelled CT3, CT4, CT5, CT6) was adjusted as 
1~:~0.8~:~1.1~:~1.3.
The ``measured'' size asymmetries are listed in the following
table; the last column gives the values expected on the basis of 
the input response factors.

\begin{center}
\begin{tabular}{|c|c|c|}
\hline
Telescopes & asymmetry & true value \\
           & $\pm 0.002$ & \\
\hline
CT3-CT4 &  0.113 &  0.111 \\
CT3-CT5 & -0.052 & -0.048 \\
CT3-CT6 & -0.128 & -0.130 \\
CT4-CT5 & -0.160 & -0.158 \\
CT4-CT6 & -0.242 & -0.238 \\
CT5-CT6 & -0.082 & -0.082 \\
\hline
\end{tabular}
\end{center}

From the first three lines, one readily determines a relative
response of the telescopes to 1~:~0.797~:~1.109~:~1.292
\footnote{Table and results rounded to three digits; 
calculations and fits used higher precision}, in agreement
with the input values to better than 1\%. With six measurements
for three independent quantities (arbitrarily defining the
response factor for CT3 as unity), one should determine optimum
calibration factors by a fit to all measurements; such a fit yields
response ratios of 1~:~0.797~:~1.104~:~1.300, with typical errors of
0.003 and a $\chi^2$ of 5.2 for 3 degrees of freedom, indicating
that small systematic effects may be present beyond the statistical
errors.

The method was then applied to actual HEGRA data, from observations
of Mkn 501 in 1997. Two independent data samples were considered,
(a) a background-subtracted gamma-ray sample obtained by 
cutting on the mean scaled width of images and on the reconstructed
direction relative to the source, and (b) a cosmic-ray sample.
The measured asymmetries are given in the following table:

\begin{center}
\begin{tabular}{|c|c|c|}
\hline
Telescopes & sample (a) & sample (b) \\
           & $\pm 0.006$          & $\pm 0.003$ \\
\hline
CT3-CT4 &  0.033 & 0.021 \\
CT3-CT5 &  0.071 & 0.073 \\
CT3-CT6 &  0.117 & 0.114 \\
CT4-CT5 &  0.037 & 0.039 \\
CT4-CT6 &  0.089 & 0.084 \\
CT5-CT6 &  0.023 & 0.045 \\
\hline
\end{tabular}
\end{center}

Fits for the telescope response factors yield
$$
\mbox{sample (a): CT3 : CT4 : CT5 : CT6 = 1 : 0.937 : 0.857 : 0.797}
$$
$$
\mbox{sample (b): CT3 : CT4 : CT5 : CT6 = 1 : 0.948 : 0.872 : 0.798}
$$
with typical errors of 0.01 for sample (a) and 0.005 for sample (b).
The results are consistent within errors; the $\chi^2$ values for the
fits are 9.7 and 7.5, again slightly worse than expected for purely
statistical errors.

In summary, the method allows an easy intercalibration of imaging
atmospheric Cherenkov telescopes in a telescope system. Both the
simulations and the comparison of two independent data samples 
shows that a precision of 1-2\% can be reached, which is ample
even for a 10\% overall energy resolution. For systems with more
than two telescopes, the equation system is overdetermined and 
consistency checks are possible, e.g. in the form of the $\chi^2$
value of an overall fit. The method can be applied to systems
with arbitrary numbers of telescopes, provided that they do not
separate into multiple distant clusters of telescopes. The absolute
calibration for the whole system can be
determined in a subsequent step, e.g., 
using cosmic-ray detection rates (see, e.g.
\cite{throughput,crcal}).

{\bf Acknowledgements.} The support of the German Ministry for
Education and Research BMBF is acknowledged. The author is 
grateful to the members of the HEGRA collaboration, who
have participated in the development, installation, and 
operation of the telescopes, and in the data precessing.

\end{document}